\documentclass{appolb}
\usepackage{graphicx}
\usepackage{amsthm}
\usepackage{amsmath}
\usepackage{amssymb}
\usepackage{esint}
\usepackage{color}
\usepackage{epsfig}
\usepackage{multirow}

\begin{document}

\eqsec  

\title{Studying and removing effects of fixed topology\thanks{Presented at ``Excited QCD 2014'', Bjelasnica Mountain, Sarajevo.}}

\author{Arthur Dromard, Christopher Czaban, Marc Wagner\address{Goethe-Universit\"at Frankfurt am Main, Institut f\"ur Theoretische Physik, Max-von-Laue-Stra{\ss}e 1, D-60438 Frankfurt am Main, Germany}}

\maketitle

\begin{abstract}
At small lattice spacing, or when using overlap fermions, lattice QCD simulations tend to become stuck in a single topological sector. Physical observables, e.g.\ hadron masses, then differ from their full QCD counterparts by $1/V$ corrections, where $V$ is the spacetime volume. These corrections can be calculated order by order using the saddle point method. We calculate all corrections proportional to $1/V^2$ and $1/V^3$ and test the resulting equations for several models: a quantum mechanical particle on a circle, the Schwinger model and SU(2) Yang-Mills theory.
\end{abstract}

\PACS{11.15.Ha, 12.38.Gc.}


\section{Introduction}

Topology freezing or fixing are important issues in quantum field theory, in particular in QCD. For example, when simulating chirally symmetric overlap quarks, the corresponding algorithms do not allow transitions between different topological sectors, i.e.\ topological charge is fixed (cf.\ e.g.\ \cite{Aoki:2008tq,Aoki:2012pma}). Also when using other quark discretizations, e.g.\ Wilson fermions, topology freezing is expected at lattice spacings $a \lesssim 0.05 \, \textrm{fm}$, which are nowadays still fine, but realistic \cite{Luscher:2011kk,Schaefer:2012tq}. There are also applications, where one might fix topology on purpose. For example, when using a mixed action setup with light overlap valence and Wilson sea quarks, approximate zero modes in the valence sector are not compensated by the sea. The consequence is an ill-behaved continuum limit \cite{Cichy:2010ta,Cichy:2012vg}. A possible solution to overcome this problem is to restrict computations to a single topological sector, either by sorting the generated gauge link configurations with respect to their topological charge or by directly employing so-called topology fixing actions (cf.\ e.g.\ \cite{Fukaya:2005cw,Bietenholz:2005rd,Bruckmann:2009cv}).

In view of these issues it is important to develop methods, which allow to obtain physically meaningful results (i.e.\ results corresponding to unfixed topology) from fixed topology simulations. The starting point for our work are calculations from the seminal papers \cite{Brower:2003yx,Aoki:2007ka}. We extend these calculations by including all terms proportional to $1/V^2$ and $1/V^3$. We apply the resulting equations to a quantum mechanical particle on a circle, to the Schwinger model and to SU(2) Yang-Mills theory and determine ``hadron masses'' at unfixed topology from fixed topology computations and simulations (for related exploratory studies in the Schwinger model and the $O(2)$ and $O(3)$ non-linear Sigma model cf.\ \cite{Bietenholz:2011ey,Bietenholz:2012sh,Bautista:2014tba}).

Part of this work has already been published \cite{Dromard:2013wja,Dromard:2014wja,Czaban:2013haa}.


\section{\label{SEC1}Hadron masses from fixed topology simulations}


\subsection{\label{SEC11}Two-point correlation functions at fixed topology}

The partition function and the two-point correlation function of a hadron creation operator $O$ at fixed topological charge $Q$ and finite spacetime volume $V$ are given by
{\small
\begin{equation}
\begin{aligned}
 & Z_{Q,V} \equiv \int DA \, D\psi \, D\bar{\psi}\, \delta_{Q,Q[A]} e^{-S_E[A,\bar{\psi},\psi]} \\
 & C_{Q,V}(t) \equiv \frac{1}{Z_{Q,V}} \int DA \, D\psi \, D\bar{\psi} \, \delta_{Q,Q[A]} O^\dagger(t) O(0) e^{-S_E[A,\bar{\psi},\psi]} .
\end{aligned}
\end{equation}
}Using a saddle point approximation the correlation function has been expanded in \cite{Brower:2003yx} according to
{\small
\begin{equation}
\label{EQN673} C_{Q,V}(t) = \alpha(0) \exp\bigg(-M_H(0) t - \frac{M^{(2)}_H(0) t}{2\mathcal{E}_2 V} \bigg(1- \frac{ Q^2}{\mathcal{E}_2 V}\bigg)\bigg) + \mathcal{O}\bigg(\frac{1}{ V^2} \bigg) ,
\end{equation}
}where $\alpha(0)$ is a constant, $M_H(\theta)$ the hadron mass at vacuum angle $\theta$, $\mathcal{E}_k \equiv e_0^{(k)}(\theta)|_{\theta=0}$ ($\mathcal{E}_2 = \chi_t$, the topological susceptibility) and $e_0$ is the vacuum energy density. In \cite{Dromard:2014wja} we have extended this calculation by including all terms proportional to $1/V^2$ and $1/V^3$,
{\small{
\begin{equation}
\label{EQN674} \begin{aligned}
 & C_{Q,V}(t) = \alpha(0) \exp\bigg(-M_H(0) t - \frac{x_2}{2\mathcal{E}_2 V} - \bigg(\frac{x_4 - 2 (\mathcal{E}_4/\mathcal{E}_2) x_2 - 2 x_2^2-4x_2Q^2}{8(\mathcal{E}_2 V)^2} \bigg) \\
 & \hspace{0.6cm} - \bigg(\frac{16 (\mathcal{E}_4/\mathcal{E}_2)^2 x_2 + x_6 - 3 (\mathcal{E}_6/\mathcal{E}_2) x_2 - 8 (\mathcal{E}_4/\mathcal{E}_2) x_4 - 12 x_2 x_4 + 18 (\mathcal{E}_4/\mathcal{E}_2) x_2^2 + 8 x_2^3}{48(\mathcal{E}_2 V)^3} \\
 & \hspace{1.2cm} - \frac{x_4 - 3 (\mathcal{E}_4/\mathcal{E}_2) x_2 - 2 x_2^2}{4(\mathcal{E}_2 V)^3} Q^2\bigg)\bigg) + \mathcal{O}\bigg(\frac{1}{(\mathcal{E}_2 V)^4} \, , \, \frac{1}{(\mathcal{E}_2 V)^4} Q^2 \, , \, \frac{1}{(\mathcal{E}_2 V)^4} Q^4\bigg) ,
\end{aligned}
\end{equation}
}where $x_n \equiv M^{(n)}_H (0) t+ \beta^{(n)}(0)$ (for the definition of $\beta^{(n)}$ cf.\ \cite{Dromard:2014wja}). The expansions (\ref{EQN673}) and (\ref{EQN674}) are rather accurate approximations, if the following conditions are fulfilled: \vspace{0.1cm}
\\\textbf{(C1)} $\phantom{xxx} 1 / \mathcal{E}_2 V \ll 1 \quad , \quad |Q| / \mathcal{E}_2 V \ll 1$. \vspace{0.1cm}
\\\textbf{(C2)} $\phantom{xxx} |x_2| = |M_H^{(2)}(0) t + \beta^{(2)}(0)| \lesssim 1$. \vspace{0.1cm}
\\\textbf{(C3)} $\phantom{xxx} m_\pi(\theta) L \gtrsim 3 \ldots 5 \gg 1$ $\ \ $ ($m_\pi$: pion mass, $L$: periodic spatial extension). \vspace{0.1cm}
\\\textbf{(C4)} $\phantom{xxx} (M_H^\ast(\theta) - M_H(\theta)) t \gg 1 \quad , \quad M_H(\theta) (T-2 t) \gg 1$. \vspace{0.1cm}

Note that the effective mass at fixed topology, defined in the usual way,
{\small
\begin{equation}
\label{eq:MQ} M^\textrm{eff}_{Q,V}(t) \equiv -\frac{1}{C_{Q,V}(t)}\frac{d C_{Q,V}(t)}{dt} ,
\end{equation}
}exhibits severe deviations from a constant behavior at large temporal separations $t$ \cite{Dromard:2014wja}, which is in contrast to ordinary quantum field theory at unfixed topology.


\subsection{\label{SEC13}Extracting hadron masses}

A straightforward method to determine physical hadron masses (i.e.\ hadron masses at unfixed topology) from fixed topology simulations is to fit either (\ref{EQN673}) or (\ref{EQN674}) to two-point correlation functions computed at fixed topology. Among the results of the fit are then the hadron mass at unfixed topology $M_H(0)$ and the topological susceptibility $\mathcal{E}_2 = \chi_t$. A similar method is to first determine hadron masses $M_{Q,V}$ at fixed topological charge $Q$ and spacetime volume $V$ and then use equations based on (\ref{EQN673}) or (\ref{EQN674}) to determine $M_H(0)$ and $\mathcal{E}_2 = \chi_t$. For a detailed discussion cf.\ \cite{Dromard:2014wja}}.


\section{\label{SEC2}A quantum mechanical particle on a circle at fixed topology}

For a first test of the methods mentioned in section~\ref{SEC13} we decided for a simple toy model, a quantum mechanical particle on a circle in a square well potential. This model shares some important features with QCD, e.g.\ the existence of topological charge and the symmetry $+\theta \leftrightarrow -\theta$. Moreover, it can be solved numerically up to arbitrary precision. We determine $M_H(0)$ (which is the energy difference between the ground state and the first excitation) and $\chi_t$ from fixed topology two-point correlation functions as outlined in section~\ref{SEC13}. We compare the $1/V$ expansion from \cite{Brower:2003yx} (eq.\ (\ref{EQN673})) and our $1/V^3$ version (eq.\ (\ref{EQN674})). We find rather accurate results for $M_H(0)$ and $\chi_t$ (cf.\ Table~\ref{TAB001}). Note that the relative errors for both $M_H(0)$ and $\chi_t$ are smaller, when using the $1/V^3$ version (\ref{EQN674}). For details cf.\ \cite{Dromard:2013wja,Dromard:2014wja}.

\begin{table}[htb]
\begin{center}
\small{
\begin{tabular}{|c|c|c|c|c|c|}
\hline 
 & expansion  & $\hat{M}_H(0)$ & error& $\hat{\chi}_t$ & error \tabularnewline
\hline 
\hline 
\multirow{2}{*}{$\frac{|Q|}{\chi_{t}V} \leq 0.5$} & (\ref{EQN674}), hep-lat/0302005 & $0.40702$ & $0.029\%$& $0.00629$ & $2.5\%$\tabularnewline
\cline{2-6} 
 & (\ref{EQN674}) &  $0.40706$ & $0.019\%$ & $0.00633$ & $1.9\%$ \tabularnewline
\hline 
\end{tabular}
}

\caption{\label{TAB001}$M_H(0)$ and $\chi_t$ from fixed topology two-point correlation functions; ``error'' denotes relative differences to the exact results $\hat{M}_{H} = 0.40714$ and $\hat{\chi}_t = 0.00645$ at unfixed topology.}
\end{center}
\end{table}


\section{The Schwinger model at fixed topology}

The Schwinger model, defined by the Lagrangian
{\small
\begin{equation}
\mathcal{L}(\psi,\bar{\psi},A_{\mu}) \equiv \bar{\psi} (\gamma_\mu (\partial_\mu + i g A_\mu) + m) \psi + \frac{1}{2} F_{\mu \nu} F_{\mu \nu} ,
\end{equation}
}also shares certain features with QCD, most prominently confinement. Furthermore, simulations are computationally inexpensive, because there are only $2$ spacetime dimensions.

We have studied the ``pion'' mass $m_\pi$ and the static quark-antiquark potential ${\mathcal V}_{q\bar{q}}$ for various separations. Results are summarized in Table~\ref{TAB002}. In the first line (``fixed top.'') results obtained from two-point correlation functions at fixed topology (as outlined in section~\ref{SEC13}) are listed. In the second line (``unfixed top.'') they are compared to results from standard lattice simulations, where gauge link configurations from all topological sectors are taken into account. One can observe agreement demonstrating that one can obtain correct and accurate physical results from fixed topology simulations. For details cf.\ \cite{Czaban:2013haa}.

\begin{table}[htb]
\begin{center}
{\small
\begin{tabular}{|c|c|c|c|c|c|c|}
\hline
 & $m_\pi a$ & ${\mathcal V}_{q\bar{q}}(1a) a$ & ${\mathcal V}_{q\bar{q}}(2a) a$ & ${\mathcal V}_{q\bar{q}}(3a) a$ & ${\mathcal V}_{q\bar{q}}(4a) a$ \\ \hline\hline
fixed top.\ &  0.2747(2) & 0.12551(4) & 0.2247(2) & 0.3005(3) & 0.3581(7)  \\ \hline
unfixed top.\ & 0.2743(3) & 0.12551(4) & 0.2247(2) & 0.3008(4) & 0.3577(9) \\ \hline
\end{tabular}
}

\caption{\label{TAB002}Comparison of results obtained from computations at fixed and at unfixed topology.}
\end{center}
\end{table}


\section{SU(2) Yang-Mills theory at fixed topology}

Currently we perform fixed topology studies of SU(2) Yang-Mills theory,
{\small
\begin{equation}
\mathcal{L}(A_\mu) \equiv \frac{1}{4} F_{\mu \nu}^a F_{\mu \nu}^a ,
\end{equation}
}which is expected to be rather similar to QCD. Again we explore the static quark-antiquark potential for various separations.

\begin{figure}[htb]
\input{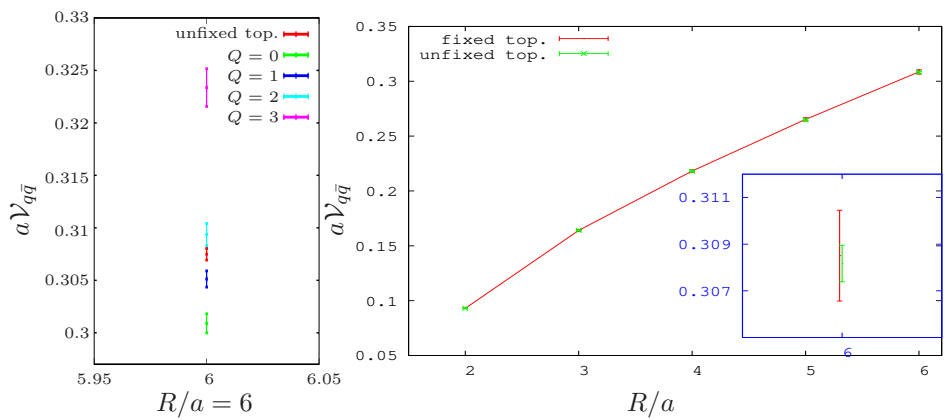}
\caption{\label{FIG342} \textbf{(left)} ${\mathcal V}_{q\bar{q}}(6a)$ for different topological sectors $Q = 0, 1, 2, 3$ for spacetime volume $V/a^4 = 16^4$. \textbf{(right)} Comparison of potential results obtained from computations at fixed and at unfixed topology.}

\end{figure}

The left plot in Fig.~\ref{FIG342} shows that there is a significant discrepancy between the potential from computations restricted to a single topological sector and corresponding results obtained at unfixed topology. The plot, therefore, underlines the necessity of a method to extract physical results from fixed topology computations.

In the right plot of Fig.~\ref{FIG342} we compare the static potential obtained from Wilson loops at fixed topology (as outlined in section~\ref{SEC13}) and from standard lattice simulations, where gauge link configurations from all topological sectors are taken into account. As for the Schwinger model, one can observe excellent agreement demonstrating again that one can obtain correct and accurate physical results from fixed topology simulations.

Details regarding our study of Yang-Mills theory at fixed topology will be published in the near future.

\section{Conclusions and outlook}

We have extended relations from the literature \cite{Brower:2003yx,Aoki:2007ka} relating two-point correlation functions at fixed topology to physical hadron masses (i.e.\ hadron masses at unfixed topology). We have successfully applied our resulting equations to various models. We plan to test the same methods for QCD in the near future, where hadron masses obtained from different topological sectors also exhibit clear differences (for an example cf.\ \cite{Galletly:2006hq}, where the pion mass has been computed in various topological charge sectors).


\section*{Acknowledgments}

We thank Wolfgang Bietenholz, Krzysztof Cichy, Dennis Dietrich, Gregorio Herdoiza, Karl Jansen and Andreas Wipf for discussions. We acknowledge support by the Emmy Noether Programme of the DFG (German Research Foundation), grant WA 3000/1-1. This work was supported in part by the Helmholtz International Center for FAIR within the framework of the LOEWE program launched by the State of Hesse.


\end{document}